\documentstyle[amsfonts,amssymb,11pt,epsfig]{article}

\makeatletter
\@addtoreset{equation}{section}

\newcommand{\R}{{\Bbb R}}
\newcommand{\C}{{\Bbb C}}

\def\SL{{\rm SL}}
\def\H{{{\bf H}^3}}

\newcommand{\be}{\begin{eqnarray}} 
\newcommand{\ee}{\end{eqnarray}}

\begin{document}

\begin{titlepage}

\thispagestyle{empty}

\title{Black Hole Thermodynamics and Riemann Surfaces \thanks{Expanded version of a talk
given in the Math Department, SUNY, Stony Brook and in the Physics Department, University
of Maryland.}}

\author{
{\bf Kirill Krasnov}\thanks{{\tt krasnov@aei.mpg.de}}
\\ \\
{\it Albert Einstein Institute, Golm/Potsdam, 14476, Germany}\\
{\it and}\\
{\it School of Mathematical Sciences, University of Nottingham}\\
{\it University Park, Nottingham, NG7 2RD, UK}}

\date{\normalsize February, 2003}
\maketitle

\begin{abstract} We use the analytic continuation procedure proposed in our earlier works to
study the thermodynamics of black holes in 2+1 dimensions. A general black
hole in 2+1 dimensions has $g$ handles hidden behind $h$ horizons. 
The result of the analytic continuation is a hyperbolic
3-manifold having the topology of a handlebody. The boundary of this handlebody is a compact Riemann surface of
genus $G=2g+h-1$. Conformal moduli of this surface encode in a simple way the physical characteristics
of the black hole. The moduli space of black holes of a given type $(g,h)$ is then 
the Schottky space at genus $G$.
The (logarithm of the) thermodynamic partition function of the hole is the K{\"a}hler potential
for the Weil-Peterson metric on the Schottky space. Bekenstein bound on the black hole 
entropy leads us to conjecture a new strong bound on this K{\"a}hler potential.
\end{abstract}

\end{titlepage}

\section{Introduction}
\label{sec:intr}

In this paper we use the analytic continuation procedure proposed in \cite{Riemann,Cont} to study
the thermodynamics of black hole (BH) solutions of 2+1 gravity with negative cosmological constant.
The simplest such black hole was described by Banados, Teitelboim, Zanelli \cite{BTZ}. Later
black holes with non-trivial internal topology were discovered \cite{Brill}. 

The standard strategy for studying BH thermodynamics is to 
analytically continue the hole spacetime. The (exponential of the) classical Einstein-Hilbert 
action evaluated on the resulting Euclidean metric is then the BH partition function. The
later can be used to obtain the BH entropy according to the usual thermodynamic formulas. Let
us remind the reader how this works for the usual Schwarzschild BH in 3+1 dimensions, see
\cite{Hawking}. The Lorentzian signature metric is given by:
\be
ds^2 = -\left(1-{2GM\over r}\right) dt^2 + {dr^2\over 1-{2GM\over r}} + r^2 d\Omega^2.
\ee
Here $M$ is the BH mass, $G$ is the Newton's constant, $d\Omega^2$ is the line
element on the unit sphere, and $t,r$ are the time and radial coordinates correspondingly.
The horizon is located at $r=r_+=2MG$. To analytically continue the BH spacetime
one sends $t\to -i\tau$. The imaginary time coordinate $\tau$ must be periodic with
period $\beta=1/T$, where $T$ is the temperature. The Euclidean metric one gets is:
\be\label{Euclid-Schw}
ds^2 = \left(1-{2GM\over r}\right) d\tau^2 + {dr^2\over 1-{2GM\over r}} + r^2 d\Omega^2.
\ee
The $r,\tau$ part of the metric describes a 2-dimensional ``plane'' with a conical
singularity at the origin $r=r_+$. The period of $\tau$ must be chosen in such a way
that there is no conical singularity. For a metric of the form:
\be\label{metric-form}
ds^2 = f(r) d\tau^2 + {dr^2\over f(r)}, \qquad f(r_+)=0
\ee
the condition that there is no conical singularity is that the circumference $\beta\sqrt{f'(r_+) \epsilon}$
of the circle $r=r_+ + \epsilon$ is equal to $2\pi$ times the proper distance 
$2\sqrt{\epsilon}/\sqrt{f'(r_+)}$ from the origin to the point  $r=r_+ + \epsilon$.
This gives for the period:
\be\label{period}
\beta = {4\pi\over f'(r_+)}.
\ee
For Schwarzschild BH $f'(r_+)=1/r_+$ and the temperature one gets from (\ref{period})
is the famous Hawking temperature:
\be
T_H = {1\over 8\pi GM}.
\ee

To obtain the BH partition function one must evaluate the Einstein-Hilbert action:
\be
I[g] = - {1\over 16\pi G} \int_{\cal M} R - {1\over 8\pi G} \int_{\partial\cal M} (K-K_0)
\ee  
on the metric (\ref{Euclid-Schw}). The first term is identically zero on shell. In the
second term $K$ is the trace of the extrinsic curvature of the boundary, and $K_0$ is
the trace of the extrinsic curvature of the boundary embedded in flat space. We have
\be
K-K_0 = - {1\over 2} f'/f,
\ee
where $f=f(r)$ is as in ({\ref{metric-form}). Then, using $f'=r_+/r^2$, computing the integral
over the boundary and sending $r$ to infinity we get:
\be\label{part-f-Schw}
I[g_{\rm cl}] = {2\pi r_+ \beta\over 8\pi G} = {4\pi r_+^2\over 4G}.
\ee
The BH partition function is then $\ln{Z} = -I[g_{\rm cl}]$. 

The standard thermodynamic relations tell us that the expectation value of the energy in the
system is given by:
\be\label{mass}
\langle E \rangle = - {\partial \ln{Z}\over \partial \beta}.
\ee
It is easy to check that $\langle E \rangle= M$, as expected. The entropy is given by:
\be\label{entropy}
S = -\beta {\partial \ln{Z}\over \partial \beta} + \ln{Z}.
\ee
The first term here equals twice the quantity (\ref{part-f-Schw}), while the second term
is minus (\ref{part-f-Schw}). Therefore, we get for the entropy:
\be
S = {4\pi r_+^2\over 4G} = {A\over 4G},
\ee
where $A$ is the horizon area. This is the famous Bekenstein-Hawking entropy.

We are going to apply the same strategy to study the thermodynamics of 2+1 dimensional
black holes. That is, we are going to analytically continue the BH spacetimes to
obtain certain spaces of Euclidean signature. To get the BH partition function we  
evaluate the gravity action on these spaces. The BH entropy is then obtained according
to the standard thermodynamic formulas.

The plan of the paper is as follows. In the next section we shall remind the
reader some facts about BH in 2+1 dimensions. In section \ref{sec:cont} we describe 
how to analytically continue the BH spacetimes.  The BH partition function is discussed in
section \ref{sec:part}. Finally, we study the BH thermodynamics in \ref{sec:thermo}.
We conclude with a short summary.

\section{Black Holes in 2+1 dimensions}
\label{sec:bh}

The material reviewed in this section is from \cite{Brill}. The references for the rotating
case are \cite{Rot,Brill-new}.

Black holes in 2+1 dimensions can have non-trivial internal topology. A general BH has
$g$ handles hidden behind $h$ horizons. 
Unlike the case with higher-dimensional black holes,
BH's in 2+1 cannot have more than one horizon per asymptotic region. Thus, the number
of horizons is also the number of asymptotic regions. 

As the gravity theory in 2+1 dimensions does not have local degrees of freedom, solutions
of vacuum Einstein equations are spacetimes of constant curvature. The case relevant to
us here is that of negative cosmological constant. Only in this case there are BH's
in the theory. Thus, we have to consider the spacetimes of constant negative curvature. They
are all locally indistinguishable from the maximally symmetric spacetime AdS${}_3$.
All (complete) spacetimes are therefore obtainable by identifications of points in AdS
acting by transformations from some discrete group $\Gamma$, subgroup of the group of
isometries.

Let us briefly remind the reader some basic facts about the Lorentzian AdS${}_3$. The spacetime is
best viewed as the interior of an infinite cylinder. The cylinder
itself is the conformal boundary $\cal I$ of the spacetime. It is
timelike, unlike the null conformal boundary of an asymptotically
flat spacetime. All light rays propagating inside AdS start and
end on $\cal I$. In this picture the constant time slices are
copies of the Poincare (unit) disk. The unit disk is isometric to
the upper half plane ${\bf U}$; we shall use both models. The
isometry group of the Lorentzian signature AdS${}_3$ is
$\SL(2,\R)\times\SL(2,\R)$. The spacetimes itself can be viewed as
the $\SL(2,\R)$ group manifold so that a point $x\in {\rm AdS}_3$ is represented
by a matrix ${\bf x}\in\SL(2,\R)$. The isometry group acts by
the left and right action: ${\bf x}\to g {\bf x} h^{-1}, g,h\in\SL(2,\R)$. 

\begin{figure}
\centerline{\hbox{\epsfig{figure=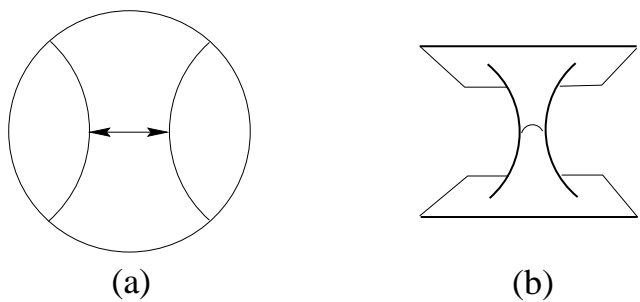,height=1.5in}}}
\caption{BTZ black hole: the geometry of the time symmetry surface.}
\label{fig:btz}
\end{figure}

Let us now return to the BH spacetimes. It is easiest to describe non-rotating BH's. In this
case there is a plane of time symmetry $X$ in the spacetime. In order to have such a plane in 
the quotient spacetime $M={\rm AdS}_3/\Gamma$, 
the discrete group $\Gamma$ that one uses to identify points 
must be such that its action fixes the $t=0$ plane in AdS. We use such a parameterization of the
$\SL(2,\R)$ group manifold that the subgroup of isometries that fixes the $t=0$ plane is
given by: ${\bf x}\to g{\bf x}g^T, g\in\SL(2,\R)$. Note that there is another ``diagonal'' action of
the $\SL(2,\R)$ given by: ${\bf x}\to g{\bf x}g^{-1}$; such transformations fix the origin of
AdS. Transformations that fix the $t=0$ plane of AdS act on this plane. The geometry of the time symmetry
plane $X$ of the quotient spacetime is therefore that of ${\bf U}/\Gamma$, where ${\bf U}$ is the
hyperbolic plane. Thus, $X$ is a Riemann surface (with holes, see below) 
uniformized by ${\bf U}$. Once the geometry of the time symmetry plane is understood
one just ``evolves'' the identifications in time to obtain the BH spacetime, see \cite{Brill}.

Let us see how this works on examples. Consider first the case of the
non-rotating BTZ BH. The corresponding discrete group is generated by a single
hyperbolic element. Its action on the $t=0$ plane can be
understood by finding the so-called fundamental region. The
fundamental region $D$ on the hyperbolic plane ${\bf U}$ for group $\Gamma$
is such that any point on ${\bf U}$ can be obtained as an image of a point
in $D$ under a transformation from $\Gamma$, and such that no
two points of $D$ (except on its boundary) are related. In the case
of $\Gamma$ generated by a single element $\gamma$ the fundamental region
is that between two geodesics on ${\bf U}$ mapped into one another
by the generator $\gamma$, see Fig.~\ref{fig:btz}(a). It is clear
that the quotient space has the topology of the
$S^1\times\R$ wormhole with two asymptotic regions,
each having the topology of $S^1$, see Fig. \ref{fig:btz}(b).
It can also be described as the geometry of a sphere with two holes.
On Fig.~\ref{fig:btz}(a) the BTZ angular coordinate runs from one
geodesics to the other. The distance between the two
geodesics measured along their common normal
is precisely the horizon circumference. Evolving the identifications
in time one obtains a spacetime of a BH with two asymptotic regions. There are two
horizons, they both intersect the time symmetry plane along the minimal length
geodesic shown in Fig. \ref{fig:btz}(b). See \cite{Brill} for more details on the
spacetime picture.

\begin{figure}
\centerline{\hbox{\epsfig{figure=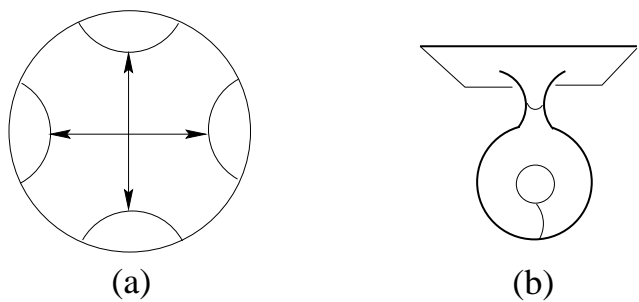,height=1.5in}}}
\bigskip
\caption{Initial slice geometry of the single asymptotic region
black hole with a torus wormhole inside the horizon}
\label{fig:wormhole}
\end{figure}

\begin{figure}
\centerline{\hbox{\epsfig{figure=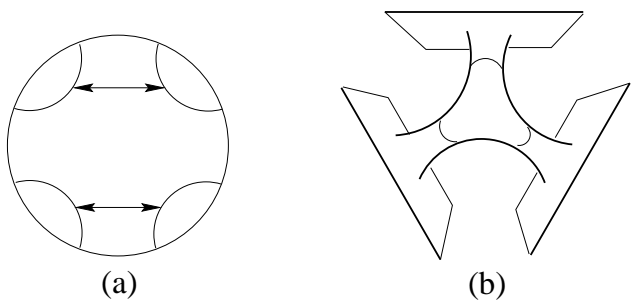,height=1.5in}}}
\caption{Initial slice geometry of the three asymptotic region black hole}
\label{fig:3bh}
\end{figure}

Let us now consider more complicated initial slice geometries.
We now consider the group $\Gamma$ to be generated by two
hyperbolic elements. For example, let the fundamental region be
the part of the unit disk between four geodesics, as in
Fig. \ref{fig:wormhole}(a). Let us identify these geodesics
cross-wise. It is straightforward to show that the resulting
geometry has only one asymptotic region, consisting of all
four parts of the infinity of the fundamental region. With
little more effort one can convince oneself that the resulting
geometry is one asymptotic region ``glued'' to a torus,
see Fig. \ref{fig:wormhole}(b). The spacetime
obtained by evolving this geometry 
is a single asymptotic region black hole, but the topology
{\it inside} the event horizon is now that of a torus.
See \cite{Brill} for more details on this spacetime.

A group generated by two elements can also be used to obtain
a three asymptotic region black hole \cite{Brill}. The fundamental
region on the $t=0$ plane is again the region bounded by four
geodesics. They are, however, now identified side-wise,
see Fig. \ref{fig:3bh}(a). One can clearly see that the
initial slice geometry has three asymptotic regions, as
in Fig. \ref{fig:3bh}(b). Evolving this, one gets a spacetime
with three asymptotic regions and corresponding event horizons.
See \cite{Brill} for more details.

Taking the group $\Gamma$ to be more complicated
one constructs a large class of spacetimes.
In particular, one can have a single asymptotic region
black hole with an arbitrary Riemann surface inside
the horizon. More generally, one can have a black hole
with any number $h$ of asymptotic regions (horizons), and with any number $g$ of
handles hidden behind the horizon(s). We shall refer to such a general BH as a hole 
of type $(g,h)$. The geometry of the time symmetry plane is that of a genus 
$g$ surface with $h$ holes. The holes are in correspondence with asymptotic regions
(horizons). We note that for $g=0$ the minimal number of holes is 2, which 
corresponds to the BTZ BH. So we have:
$h\geq 2, g=0$. For $g\geq 1, h\geq 1$.

Let us now consider general, rotating black holes. The idea of \cite{Rot} was
to unravel the spacetime structure by considering the action of
the discrete group at the boundary cylinder, instead of thinking about the
action at the $t=0$ plane that is no longer preserved by identifications.
One gets a rather effective description, which, for example, allows one to
calculate horizon angular velocities in terms of traces of group elements
generating identifications. The following formula was obtained in \cite{Cont}
using methods of \cite{Rot}. To get the angular velocity of a horizon, one
must find a group element that generates isometries of the corresponding
asymptotic region. The angular velocity is then obtained as:
\begin{equation}\label{J*}
\Omega = {{\rm Arccosh}\left({1\over 2}{\rm Tr}\,\gamma^L \right)-
{\rm Arccosh}\left({1\over 2}{\rm Tr}\,\gamma^R \right)\over
{\rm Arccosh}\left({1\over 2}{\rm Tr}\,\gamma^L \right)+
{\rm Arccosh}\left({1\over 2}{\rm Tr}\,\gamma^R \right)}.
\end{equation}
Here $\gamma^L, \gamma^R$ are the left and right parts of an isometry
$\gamma\in\SL(2,\R)\times\SL(2,\R)$. For example, one makes the wormhole of
Fig. \ref{fig:wormhole} rotating by taking two generators $\gamma_1, \gamma_2$
whose left and right parts are not equal. A generator
of isometries of the single asymptotic region of this wormhole is
then the commutator: $\gamma = [\gamma_1,\gamma_2]$. Substituting its
left and right parts into the formula (\ref{J*}) one gets the horizon
angular velocity. Similarly one can compute the angular velocity of any horizon
for a BH of a general type. 

We also note that there is a similar formula for the
horizon size:
\be\label{size*}
2\pi r_+ = {\rm Arccosh}\left({1\over 2}{\rm Tr}\,\gamma^L \right)+
{\rm Arccosh}\left({1\over 2}{\rm Tr}\,\gamma^R \right).
\ee
Here $\gamma^L, \gamma^R$ are also the left and right parts of a generator of
isometries of the asymptotic region whose horizon size is being computed.
Formulas (\ref{J*}), (\ref{size*}) allow us to calculate all of the horizon properties in terms
of the corresponding group elements. An analytically continued version of these
formulas will play an important role in what follows.

\section{Analytic Continuation}
\label{sec:cont}

The content of this section is from \cite{Riemann,Cont}.

Let us now turn to the procedure of analytic continuation. We first describe this
procedure for a non-rotating BH. The
basic idea is, instead of analytically continuing the metric in
some time coordinate, produce a space $\cal M$ by identifying points in the
Euclidean AdS${}_3$ {\it using the same group} $\Gamma$. That is
${\cal M}=\H/\Gamma$. We recall
that the group of isometries of the Euclidean AdS${}_3$
(=hyperbolic space $\H$) is $\SL(2,\C)$. However, $\SL(2,\R)$ is naturally a subgroup of
$\SL(2,\C)$, thus $\Gamma$ acts on $\H$ and this action can be
used to obtain a quotient space. To see what this quotient space
is let us give another, equivalent description of the continuation
map. Let us take a section of the Poincare ball (a model for $\H$)
by a plane passing through the center of the ball. The
intersection of the ball with the plane is a unit disk ${\bf U}$.
Let us call this $t=0$ plane and do on it the same identifications
as we do on the time symmetry plane of the spacetime to be
analytically continued. Let us then ``evolve'' these
identifications, but now in the Euclidean time. To do this one
just constructs geodesic surfaces intersecting the $t=0$ plane
orthogonally along the geodesics bounding the fundamental region.
The geodesic surfaces in $\H$ are hemispheres; they are to be
identified.

Let us see how this works for the simplest case of the BTZ black hole.
Thus, we require that the geometry of the $t=0$ slice of the
unit ball is the same as the geometry of the
$t=0$ slice of BTZ black hole, see Fig. \ref{fig:btz}. We then
have to build geodesic surfaces
above and below the two geodesics on the $t=0$ plane, see
Fig. \ref{fig:btz-eucl}(a). The Euclidean
BTZ black hole is then the region between these hemispheres;
the hemispheres themselves are identified. It is clear that
the space obtained is a solid torus, its conformal boundary
being a torus. It is often more convenient to
work with another model for the same space,
that using the upper half space. The interior of the Poincare
ball can be isometrically mapped into the upper half-space.
The boundary sphere goes under this map into the $x-y$ plane.
In the case of $\Gamma$ generated by a single generator one
can always put its fixed points to $0,\infty$, so that
the picture of the Euclidean BTZ BH becomes that in
Fig.~\ref{fig:btz-eucl}(b). It is important that
using our procedure we have arrived at the same space
as is the one obtained by the usual analytic continuation
in the time coordinate, see \cite{Carlip-T}.

\begin{figure}
\centerline{\hbox{\epsfig{figure=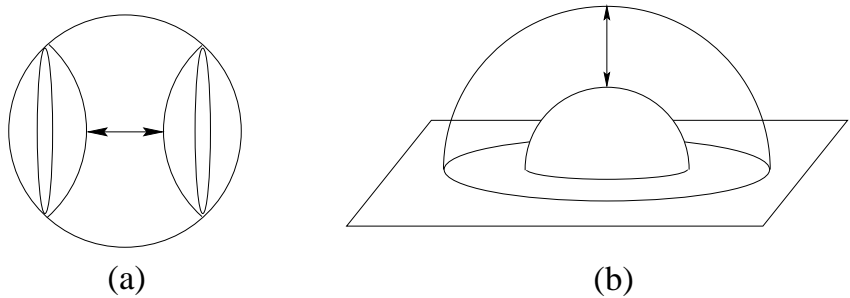,height=1.5in}}}
\caption{Euclidean BTZ black hole}
\label{fig:btz-eucl}
\end{figure}

Let us consider another example. We now want to construct the
Euclidean version of the single asymptotic region black hole
with a torus inside the horizon. The procedure is the same:
we require a slice of the unit ball to have the same geometry
as the $t=0$ slice of the black hole. This gives us four
hemispheres inside the unit ball; the Euclidean space is the
region between them and they are to be identified cross-wise,
see Fig. \ref{fig:wormhole-eucl}(a). One sees that the
Euclidean space is a solid 2-handled sphere. One can again
map the whole configuration into the upper half-space,
see Fig. \ref{fig:wormhole-eucl}(b).

\begin{figure}
\centerline{\hbox{\epsfig{figure=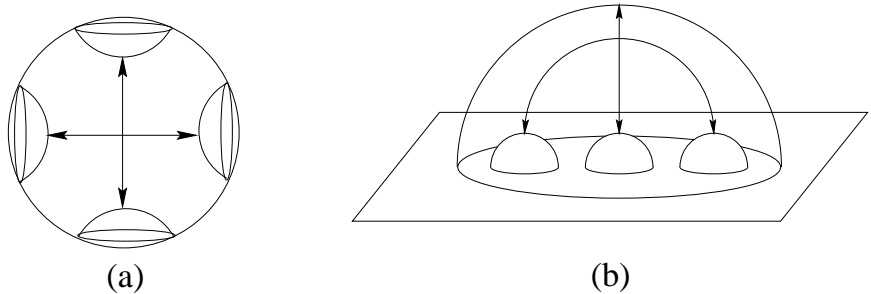,height=1.5in}}}
\caption{Euclidean single asymptotic region black hole with a torus inside}
\label{fig:wormhole-eucl}
\end{figure}

One can do a similar analysis for the three asymptotic region
black hole (one also gets a solid two-handled sphere),
and for any other of the non-rotating black holes
of \cite{Brill}. In all cases the pattern is the same: one
requires the $t=0$ slice geometry to be the same also in the
Euclidean case, and this determines the Euclidean geometry
completely. The Euclidean spaces one gets are handlebodies.

We would now like to understand in more detail a relation between
the geometry of the time symmetry plane and the conformal boundary $\partial{\cal M}$
of the Euclidean space. As we shall explain, the later is
the so-called Schottky double of the former. Given a Riemann surface $X$,
closed or with a boundary, its Schottky double is another Riemann
surface $\tilde{X}$, not necessarily connected, out
of which the original surface can be obtained by
identifications: $X=\tilde{X}/\sigma$. Here $\sigma$
is an anti-holomorphic map of $\tilde{X}$ into itself.
For a surface $X$ without a boundary the Schottky double
$\tilde{X}$ is given by two disconnected copies
of $X$, with orientation of the second copy reversed. For a surface with a boundary one takes
two copies of $X$ and glues them along the boundary to obtain
a connected surface. The anti-holomorphic map $\sigma$ fixes
the pre-image of the boundary of $X$ on $\tilde{X}$.

It is not hard to convince oneself that the geometry of the
boundary $\partial{\cal M}$ of the Euclidean space ${\cal M}=\H/\Gamma$ obtained via our analytic
continuation prescription is that of the Schottky double of the
$t=0$ geometry $X$. This is related to the fact that the Riemann
surface one obtains is uniformized by the complex plane, which is
the so-called uniformization by Schottky groups. Let us first
recall some basic information about the Schottky groups, see,
e.g., \cite{Hyperb} as a reference. A Schottky group $\Sigma$ is a
discrete subgroup of $\SL(2,\C)$, freely (that is, no relations)
generated by a number $g$ of loxodromic (that is ${\rm Tr}(L_i)
\notin [0,2]$) generators $L_1,\ldots,L_g\in\SL(2,\C)$. The
Schottky group $\Sigma$ acts by conformal transformations on the
complex plane $\C$. Let us denote by $\cal C$ the complement of
the set of fixed points of this action. As is not hard to convince
oneself, the quotient ${\cal C}/\Sigma$ is a compact genus $g$
Riemann surface. A Riemann surface obtained from the complex plane
by identifications from a Schottky group is called uniformized via
Schottky. This is a uniformization different from the usual
Fuchsian one that uses the hyperbolic plane. We have already
encountered surfaces uniformized by Schottky groups. The
boundaries of our Euclidean spaces were obtained exactly this way.
It is only that we considered Schottky groups that are real, that
is, subgroups of $\SL(2,\R)$. This is related to the fact that we
have so far only considered non-rotating spacetimes. As we explain
later, inclusion of rotation would amount to considering general
Schottky groups.

\begin{figure}
\centerline{\hbox{\epsfig{figure=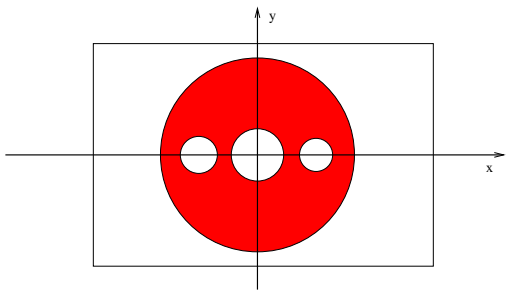,height=1.5in}}}
\caption{The fundamental domain for the Schottky uniformization
of a genus 2 surface. The complex plane here is the conformal boundary of the
hyperbolic space. Adding above the plane 4 hemi-spheres based on these circles one gets the
configuration depicted in Fig. \ref{fig:wormhole-eucl}.}
\label{fig:Schottky}
\end{figure}

Let us illustrate all this on an example. Consider the single
asymptotic region wormhole. The $t=0$ plane geometry is obtained
as a quotient with respect to a group $\Gamma\subset\SL(2,\R)$
generated by two elements. The same group, thought of as a
subgroup of $\SL(2,\C)$ acts in $\H$. In particular, it acts on
the boundary of $\H$, that is the complex plane, by fractional
linear transformations. The boundary of the Euclidean space is
then obtained as the quotient of the complex plane by this action.
The fundamental region for this action is shown in
Fig.~\ref{fig:Schottky}. Since all generators are in $\SL(2,\R)$,
their fixed points are located on the real axes, and so are the
centers of the circles bounding the fundamental region. Removing
the circles one gets a sphere with four holes. Identifying their
boundaries one gets a genus 2 surface --our Euclidean boundary.
Extending the identifications into the bulk of the hyperbolic space
one gets the situation depicted in Fig. \ref{fig:wormhole-eucl}.
The $t=0$ plane geometry is embedded into this figure as the
$z-x$ plane, that is as a plane orthogonal to the boundary of $\H$
and intersecting it along the real axis. Identifications induced on
this $z-x$ plane are precisely those needed to get $X$. It is now
clear that the geometry of the Euclidean boundary $\partial{\cal M}$ is precisely
that of the double of $X$. Let us note that the two copies of $X$
needed to obtain the Euclidean boundary are exactly the ``same'' (but have opposite orientation).
Indeed, the configuration of circles in Fig.~\ref{fig:Schottky} is
invariant under the reflection on the real line. This reflection
is exactly the anti-holomorphic map $\sigma$ that is part of the
definition of the Schottky double. 

Having explained why the boundary of the Euclidean space is
the Schottky double of the initial slice geometry, let use this
fact to obtain a simple relation between the number of asymptotic
regions $h$, the number of handles $g$ behind the horizon, and the genus
of the Euclidean boundary $G$. As it is easy to see:
\begin{equation}
G = 2g + h - 1.
\end{equation}

Let us now describe the spaces obtained as the analytic continuation of rotating
BH spacetimes. The main idea is \cite{Cont} to analytically continue
the identifications in AdS${}_3$ into isometries of the
hyperbolic space. This is achieved by analytically continuing the coordinates
of the fixed points of the Lorentzian isometries on the boundary cylinder. 
The result of this analytic continuation is
a complex group $\Sigma\in\SL(2,\C)$, so that the Euclidean space that corresponds to
a rotating BH spacetime is ${\cal M}=\H/\Sigma$. 

To understand what class of groups $\Sigma$ may arise we need to understand the meaning
of the deformation due to rotation. As is explained in detail in \cite{Cont}, the
complex group $\Sigma$ is best thought of as a certain deformation of the group of
the non-rotating spacetime $\Gamma_{\rm non-rot}\in\SL(2,\R)$. The deformation in
question is the so-called quasi-conformal deformation due to a Fenchel-Nielsen twist.

\begin{figure}
\centerline{\hbox{\epsfig{figure=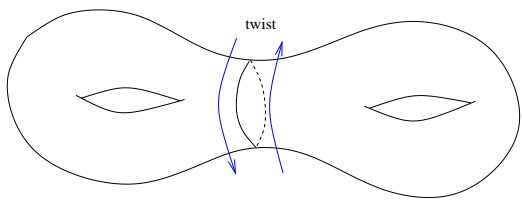,height=1in}}}
\caption{Turning on rotation in some asymptotic region is equivalent to making a Fenchel-Nielsen twist
on the corresponding geodesic on $\partial {\cal M}$. An example of the single asymptotic region
torus wormhole is shown.}
\label{fig:twist}
\end{figure}

The Fenchel-Nielsen twist is a way to change the conformal structure on a Riemann surface.
One selects a simple geodesic, cuts the surface along this geodesic, rotates the sides of the
cut with respect to each other, and glues back. One obtains a Riemann surface with a different
conformal structure. One can describe this new conformal structure either by Fuchsian
or quasi-Fuchsian groups. The quasi-Fuchsian description was first given by Wolpert
\cite{Wol-1}. In this description one starts with a Fuchsian group $\Gamma$ corresponding to
the original surface. The twist along some geodesic corresponds to a particular 
quasi-conformal $f^\tau$ map of the upper half-plane into the interior of a Jordan curve. The
image of the real axis is the Jordan curve itself. The resulting deformed group 
$\Gamma^\tau = f^\tau \circ \Gamma \circ (f^\tau)^{-1}$ is a quasi-Fuchsian group, 
subgroup of $\SL(2,\C)$. 

As was explained in \cite{Cont} the description in terms of quasi-Fuchsian groups is the one
relevant in our case. One starts with a non-rotating spacetime. Its analytic continuation
is the manifold $ {\cal M}=\H/\Gamma$, whose boundary $\partial {\cal M}$ is a Riemann surface:
the Schottky double of the time symmetry plane. There is a set of simple closed geodesics on
$\partial {\cal M}$ that correspond to asymptotic regions (horizons). Turning on rotation
in one of the asymptotic regions is equivalent to making a Fenchel-Nielsen twist along 
the corresponding horizon geodesic on $\partial {\cal M}$, see Fig. \ref{fig:twist}. 
As the result of the twist
one gets a quasi-Fuchsian group $\Sigma\in\SL(2,\C)$. The size and angular velocity of all
horizons can be easily expressed in terms of traces of certain group elements from 
$\Sigma$. Taking any group element $A\in\Sigma$ whose axis projects on a horizon geodesics on $\partial {\cal M}$
one has:
\be\label{moduli}
{1\over 2} {\rm Tr} A = \cosh\left({\pi(r_+ + i|r_-|)\over l}\right),
\ee
where $r_+, r_-$ are the outer and inner horizon radii correspondingly. The mass and angular momentum
of the horizon can be expressed in terms of $r_+, r_-$ by simple formulas, see below. Thus, the
analytic continuation of the rotating spacetime is the manifold ${\cal M}=\H/\Sigma$, where
$\Sigma$ is a complex group. The moduli of the Riemann surface $\partial{\cal M}$, such as
(\ref{moduli}) describe physical properties of the BH. 

Let us now discuss what is the most general group $\Sigma\in\SL(2,\C)$ that corresponds to 
a BH spacetime. The boundary $\partial {\cal M}$ of the analytically continued BH spacetime
is a quotient ${\cal C}/\Sigma$, where ${\cal C}$ is the domain of discontinuity for
the action of $\Sigma$ on $\C$. The surface $\partial{\cal M}$ is the double of a Riemann
surface $X$ with $g$ handles and $h$ holes. The number of moduli describing the conformal
geometry of $X$ is $6g+3h-6$, out of which $h$ moduli describe the sizes of the holes.
The most general $\partial {\cal M}$ is obtained by taking two surfaces $X$ with different
conformal geometry, and gluing them across the holes. The only condition is that the
sizes of holes in the two copies of $X$ must be the same. However, there is an additional
parameter --twist-- for each hole. Thus, overall, one gets $2 (6g+3h-6)$ real parameters. This equals
$6G-6$, the dimension of the moduli space at genus $G$, where $G$ is the genus of $\partial{\cal M}$.
Thus, the dimension of the moduli space of $\Sigma$ that correspond to BH spacetimes is just
the dimension of the Teichm{\"u}ller space at genus $G$.

To describe such $\Sigma$'s more explicitly we need a version of the Bers' simultaneous uniformization
theorem. In its classical version this theorem states that, given two Riemann surfaces $X, Y$ of the
same genus and opposite orientation, there exists a quasi-Fuchsian group $\Sigma\in\SL(2,\C)$ that
simultaneously uniformizes them. In other words the surface ${\cal C}/\Sigma$, where as
usual ${\cal C}$ is the domain of discontinuity for the action of $\Sigma$ on $\C$, is the
disconnected sum $X\cup Y$. The theorem is proved by showing that there exist a quasi-conformal
map $f^1$, which is conformal in the lower half-plane, such that $f^1 \circ \Sigma \circ (f^1)^{-1}$
equals to the Fuchsian group $\Gamma_X$ uniformizing $X$. Similarly, there exists a quasi-conformal
map $f^2$, conformal in the upper half-plane, such that $f^2 \circ \Sigma \circ (f^2)^{-1}$ is the
Fuchsian group $\Gamma_Y$ uniformizing $Y$.

What we need is a version of this theorem for simultaneous uniformization of two surfaces with holes.
Let $X, Y$ be two surfaces of type $(g,h)$ of opposite orientation. The statement that we need is that
given $X, Y$ and a set of $h$ numbers --twists-- that tell us how to glue $X$ to $Y$ across the holes,
there exists a complex group $\Sigma$ that uniformizes the surface obtained by gluing $X$ to $Y$. Similarly
to the compact $X, Y$ case one expects that there are two quasi-conformal maps, conformal in the lower and upper
half-planes correspondingly, such that $\Sigma$ ``untwisted'' by these maps gives the Fuchsian groups
$\Gamma_X, \Gamma_Y$ of the second kind uniformizing $X, Y$. This
would be a natural analog of the classical simultaneous uniformization. Unfortunately,
we are not aware of a statement to this effect in the literature. It would be of interest to generalize
simultaneous uniformization to the case of surfaces with holes.

The groups $\Sigma$ that arise as a result of such simultaneous uniformization of two surfaces with
holes $X, Y$ are the most general groups that correspond to BH spacetimes. The physical properties
of BH's can be read off from the group directly using the formula (\ref{moduli}). The groups arising
are finitely generated, free, purely loxodromic Kleinian groups with one component. 
According to a theorem due to Maskit \cite{Maskit} these are exactly the Schottky groups. Thus, our $\Sigma$ are
Schottky groups, and the moduli space of analytic continuations of BH spacetimes is the Schottky space.
We are now ready to calculate the thermodynamic partition function for our BH's. We do this by evaluating
the Einstein-Hilbert action on the Euclidean 3-manifolds ${\cal M}$ --analytic continuation of
BH spacetimes. The partition function will be a certain function on the Schottky space.

\section{Partition Function}
\label{sec:part}

This section is based on \cite{Riemann,Takht-Teo}. The Einstein-Hilbert action to be evaluated on 
spaces ${\cal M}$ is given by:
\be\label{action}
I[g] = - {1\over 16\pi G} \int_{\cal M} \sqrt{g} (R-2\Lambda) - {1\over 8\pi G}
\int_{\partial {\cal M}} \sqrt{\gamma} K + {l\over 8\pi G} \int_{\partial {\cal M}} \sqrt{\gamma}.
\ee
Here $\gamma$ is the restriction of the metric $g$ on ${\cal M}$ to the boundary $\partial{\cal M}$.
The last boundary term proportional to the boundary area is necessary to regularize the action.
To evaluate the action on ${\cal M}=\H/\Sigma$ one must select a regularizing family of surfaces.
There is a canonical family in $\H$ with the following properties: (i) it is compatible with
identifications one makes to get ${\cal M}$; (ii) each surface has constant negative curvature. Explicitly,
the surfaces are given by:
\begin{eqnarray}\label{coord}
\xi &=& {\rho\, e^{-\varphi/2}\over 1+ 
{1\over 4}\rho^2 e^{-\varphi} |\varphi_w|^2}, \\
\nonumber
y &=& w + {\varphi_{\bar{w}}\over 2} 
{\rho^2 e^{-\varphi}\over 1+ {1\over 4}\rho^2 e^{-\varphi} |\varphi_w|^2}.
\end{eqnarray}
Here $\xi, y$ are the usual coordinates in $\H$ in the upper half-space model. Fixing
$\rho=const$ one gets surfaces in question. Coordinates $w,\bar{w}$ (and $\rho$) 
are Gaussian coordinates based on these surfaces. 

The key quantity in (\ref{coord}) is the {\it canonical Liouville
field} $\varphi$. It is a (real) function of the complex coordinate
$w\in{\cal C} : \varphi=\varphi(w,\bar{w})$. It depends in a certain way
on the Schottky group $\Sigma$: it satisfies the Liouville
equation on the Schottky domain 
${\cal C}$ (domain of discontinuity of the action of $\Sigma$ on $\C$)
and has the following transformation property:
\begin{equation}\label{trans-prop}
\varphi(L w) = \varphi(w)-\ln{|L'|^2}.
\end{equation}
The Liouville field can be constructed if the map between the
Schottky and Fuchsian uniformization domains is known, see, e.g., \cite{Takht}
for more details. 

Evaluation of (\ref{action}) on ${\cal M}$ reduces to computation of the volume of the
part of the fundamental region in $\H$ that lies above a surface $\rho=\epsilon$. One then
subtracts a multiple of the area of this surface, removes a simple logarithmic divergence proportional
to the Euler characteristic $\chi=2-2G$, and takes the limit $\epsilon\to 0$. The result of
this computation is the Liouville action on $\partial{\cal M}$, as defined in
\cite{Takht}, evaluated on the canonical Liouville field $\varphi$:
\be\label{part-f}
I[{\cal M}] = -{l\over 4G} I_{\rm Liouv}[\varphi].
\ee
See \cite{Riemann,Takht-Teo} for more details on this computation. As it was shown
in \cite{Takht}, the on-shell Liouville action appearing in (\ref{part-f}) is
the K{\"a}hler potential for the Weil-Peterson metric on the Schottky space. In other words,
the (logarithm of the) thermodynamic partition function of a BH, as a function of BH physical parameters
(encoded by moduli of $\partial{\cal M}$) is the K{\"a}hler potential on the BH moduli space. Knowing the
BH partition function we are ready to study the BH thermodynamics.

\section{Thermodynamics}
\label{sec:thermo}

The content of this section is new. Our goal will be to obtain the BH entropy. We use the 
standard thermodynamics relations. What we need is analogs of
relations (\ref{mass}), (\ref{entropy}). It is instructive to consider the case of the BTZ BH first.
Our treatment of the BTZ BH is reminiscent of \cite{Carlip-T}.

Let us first introduce convenient notations. Define:
\be
\alpha = {p\over 2b}, \qquad p = {r_+ + i|r_-|\over l}, 
\ee
and $b$ is defined via $1/b^2 = l/4G$.  We remind the reader that $l$ is the radius of curvature
of AdS, defined as $l=1/\sqrt{-\Lambda}$, where $\Lambda$ is the cosmological constant. We set
$\hbar=1$, so that $1/b^2$ is essentially the ratio of radius of curvature of AdS to the Planck length.
Semi-classical regime of small curvatures corresponds to $b\to 0$. Define:
\be
\Delta-\Delta_0 = \alpha^2.
\ee
All these quantities have interpretation in terms of Liouville theory: $b$ is the Liouville coupling
constant and $\Delta-\Delta_0$ is the conformal dimension of a ``primary state'' $\alpha$, minus
the conformal dimensions of the lowest lying state. This
relation to Liouville theory won't play any role in this paper, we mention it only to provide an
explanation for the relations above. The mass and angular momentum of the hole can now be expressed
entirely in terms of the quantity $\alpha$, more precisely in terms of the conformal dimension.
One has:
\be\label{mass-BTZ}
M l  = (\Delta-\Delta_0) + \overline{(\Delta-\Delta_0)}, \\ \nonumber
J/ 8G = (\Delta-\Delta_0) - \overline{(\Delta-\Delta_0)}.
\ee
Here $M, J$ are the BH mass and angular momentum. Thus, we see that $M, J$ are completely determined
by the parameter $\alpha$. It will be much more convenient to use the complex parameter 
$\alpha$ as a physical parameter describing the hole.

\begin{figure}
\centerline{\hbox{\epsfig{figure=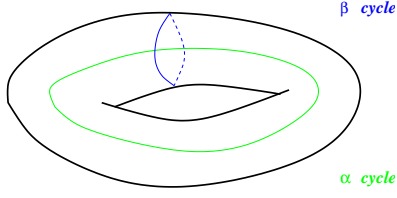,height=1in}}}
\caption{The alpha- and beta-cycles on the boundary of the Euclidean BTZ BH. The alpha-cycle is non-contractible
inside the hole. The beta-cycle is contractible, and is in the Euclidean time direction.}
\label{fig:a-b}
\end{figure}

To introduce an analog of the relation (\ref{mass}) we, following \cite{Carlip-T}, consider an
``off-shell'' BTZ BH. In the Euclidean BTZ BH the alpa-cycle on the boundary is not-contractible
inside the space, see Fig. \ref{fig:a-b}. The corresponding $\SL(2,\C)$ holonomy is non-trivial and is just the
generator of $\Sigma\subset\SL(2,\C)$, the group that we use to obtain the space ${\cal M}=\H/\Sigma$.
The other, conjugate generator of $\pi_1(\partial {\cal M})$, the beta-cycle is contractible
inside ${\cal M}$. The corresponding $\SL(2,\C)$ holonomy is trivial. The off-shell BTZ BH
is obtained by allowing the beta-cycle holonomy to be non-trivial. Thus, one takes:
\be
{1\over 2} {\rm Tr} A_\beta = \cos{\pi(\phi+i\chi)},
\ee
and introduces a variable:
\be
\beta = {1\over 2b}(\phi+i\chi).
\ee
The on-shell BTZ BH is obtained by setting $\phi=1, \chi=0$, in other words $\beta=1/2b$. 

We should now calculate the partition function for the off-shell BTZ BH. It is a straightforward
exercise to do this. First, we note that for the on-shell BTZ the Euclidean action one gets is:
\be\label{on-shell}
I_{\rm BTZ} = - {2\pi r_+\over 8G}.
\ee
It equals to minus half of the corresponding Bekenstein-Hawking entropy. This result can
be obtained by evaluating the volume between two hemi-spheres in $\H$, regularized by the
family of surfaces (\ref{coord}), which in the BTZ case is simply a family of cones with the tip
at the origin. See, e.g., \cite{Riemann} for this calculation. Modification for an off-shell
BTZ is simple. Note that when $\phi\not=1, \chi=0$, one has to cut out a wedge of angle $2\pi\phi$ 
out of $\H$. This means that the factor of $2\pi$ in (\ref{on-shell}) will be replaced
by $2\pi\phi$. One can similarly calculate the action for the general case $\chi\not=0$. The
result when written in terms of $\alpha, \beta$ complex variables is:
\be
\ln{Z} = 2\pi \left(\alpha\beta + \overline{\alpha\beta}\right).
\ee 
We wrote the result in terms of the partition function: $\ln{Z}=-I$.
Thus, we have the following relation between the $\alpha$ and $\beta$ parameters:
\be\label{alpha}
\alpha = {1\over 2\pi} {\partial\ln{Z}\over \partial \beta}.
\ee
This is the desired analog of the relation (\ref{mass}) for BTZ BH. Of course, one could have also
take a derivative of the $\ln{Z}$ with respect to the usual inverse temperature, as it was done
in (\ref{mass}). What one would obtain is the BTZ mass (\ref{mass-BTZ}). It is, however,
more convenient to work with the parameters $\alpha, \beta$ directly related to the holonomies
of the BH. 

The entropy can now be obtained using:
\be\label{entropy-btz}
S = \left(\beta {\partial\ln{Z}\over\partial\beta} + \bar{\beta}{\partial\ln{Z}\over\partial\bar{\beta}}
\right) + \ln{Z}= 2\pi \left(\alpha\beta + \overline{\alpha\beta}\right) + \ln{Z}.
\ee
Here we have used (\ref{alpha}) to write the second equality. To obtain the on-shell BTZ BH 
entropy we must set $\beta = 1/2b$ in the above expression. One obtains:
\be
S_{\rm BTZ} = {2\pi r_+\over 4G}.
\ee
This is the Bekenstein-Hawking entropy equal to the horizon length devided by $4G$.

Using this example as a guide, let us now consider a general BH of type $(g,h)$. Its analytic
continuation is the 3-manifold ${\cal M}$, whose boundary $\partial{\cal M}$ is a Riemann surface of genus
$G=2g+h-1$. The boundary is marked with a set of $h$ simple closed geodesics: horizon geodesics.
We shall refer to these geodesics as alpha-cycles. In addition, there is a set of beta-cycles
on $\partial {\cal M}$. These are such that the $\SL(2,\C)$ holonomy along them is trivial.
They are contractible inside ${\cal M}$. Each beta-cycle is intersected by exactly two alpha-cycles,
see Fig. \ref{fig:a-b-3} for a drawing of the cycles for a 3-asymptotic region BH.

\begin{figure}
\centerline{\hbox{\epsfig{figure=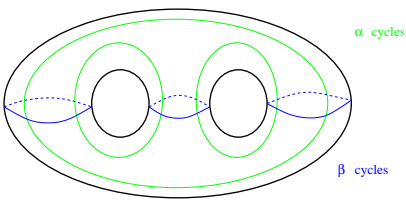,height=1in}}}
\caption{Alpha- and beta-cycles on the boundary of the Euclidean 3-asymptotic region BH.}
\label{fig:a-b-3}
\end{figure}

The first step towards general BH thermodynamics would be to evaluate the partition function for 
a general off-shell BH, in which both alpha- and beta-cycles are non-trivial. This is equivalent
to a problem of evaluating the action (\ref{action}) on a 3-manifold with a set of conical 
singularities inside. The moduli space of such manifolds is parameterized by homomorphisms of
$\pi_1(\partial {\cal M})$ into $\SL(2,\C)$, modulo conjugations by $\SL(2,\C)$. 
 This moduli space has the real dimension $12 G-12$. The on-shell Euclidean BH's form a
submanifold of real dimension $6G-6$ in it. The moduli space of off-shell BH's is naturally
a symplectic manifold. The symplectic form is that of CS theory on $\partial{\cal M}$. Some properties
of this phase space were studued in \cite{Teich}, where also a relation to the most general solution
of 2+1 gravity with negative cosmological constant was established. However, those details will not
be of importance for us here. 

Let us denote by $I_{\rm off-shell}$ the result of the evaluation of the
action (\ref{action}) on an off-shell BH. Introducing the parameters $\alpha, \beta$ for
all alpha- and beta-cycles, the off shell action becomes a function of these parameters. The
on-shell result   (\ref{part-f}) suggests that the general off-shell action is the K{\"a}hler
potential on the general moduli space 
\be\label{off-shell-moduli}
{\rm Hom}\left (\pi_1(\partial{\cal M}),\SL(2,\C)\right)/\SL(2,\C).
\ee
As $\alpha, \beta$ are the canonically conjugate quantities on this moduli space, this would imply
that
\be\label{1}
\alpha = {1\over 2\pi} {\partial \ln{Z}\over \partial\beta},
\ee
exactly like what we had in the BTZ BH case. Note, however, that the function $\ln{Z}$ in the
general case is much more complicated and is not given by a simple product of $\alpha\beta$ like
in the BTZ case. In fact, it is an outstanding problem to compute $\ln{Z}$ in some explicit fashion.

Assuming that our conjecture about the off-shell $\ln{Z}$ being the K{\"a}hler potential on 
(\ref{off-shell-moduli}) is correct, we can compute the BH entropy. It is given by an
analog of the formula (\ref{entropy-btz}):
\be\label{entropy-gen}
S = \sum \left(\beta {\partial\ln{Z}\over\partial\beta} + \bar{\beta}{\partial\ln{Z}\over\partial\bar{\beta}}
\right) + \ln{Z}= 2\pi \left(\alpha\beta + \overline{\alpha\beta}\right) + \ln{Z}.
\ee
Here the sum is taken over all the horizon geodesics, and we have used (\ref{1}) to write the
second equality. We are interested in the on-shell BH entropy. To get this, we need to set
all $\beta=1/2b$ in  (\ref{entropy-gen}). We get:
\be\label{entropy-on-shell}
S = \sum {2\pi r_+\over 8G} + \ln{Z}.
\ee
Thus, the on-shell entropy is given by the sum of horizon circumferences over $8G$ plus the
logarithm of the partition function, proportional to the K{\"a}hler potential on the Schottky space.
This is our main formula for the entropy of a general BH. We see that the entropy is not given
by a simple sum of length of the horizons. There is a contribution depending non-trivially on
the other moduli. It would be of interest to study this dependence in detail.

We could have derived (\ref{entropy-on-shell}) without any recourse to the off-shell BH. Indeed, we could
have written for the entropy:
\be\label{2}
S = \sum \left( \beta M + \Phi J\right) + \ln{Z},
\ee
where the sum is taken over all horizons, and $M, J$ are the usual mass and angular momentum in an
asymptotic region, and $\beta, \Phi$ are the thermodynamically conjugate quantities. Since every asymptotic
region is indistinguishable from that of the BTZ BH, the quantity in brackets in (\ref{2}) 
is equal to $2\pi r_+/8G$, which gives (\ref{entropy-on-shell}). 

Let us now note that physical considerations, namely the Bekenstein bound for the entropy of a system,
suggest that the entropy of a general BH of type $(g,h)$ is bounded by the sum of horizon circumferences over
$4G$:
\be\label{bound}
S \leq \sum {2\pi r_+\over 4G}.
\ee
In view of (\ref{entropy-on-shell}) the bound (\ref{bound}) becomes a bound on the K{\"a}hler potential
that can be written as:
\be\label{bound*}
I_{\rm Liouv}[\Sigma] \leq \sum \log{|m|}.
\ee
Here $I_{\rm Liouv}[\Sigma]$ is the K{\"a}hler potential on the Schottky space that appears in (\ref{part-f}), 
and is equal to the Liouville action evaluated on the canonical Liouville field; $\Sigma$ is a point
in the Schottky space. The sum is taken over all horizon geodesics, and $m$ is the mutliplier of
the corresponding transformation, that is the modulus greater than 1 eigenvalue of the corresponding
$\SL(2,\C)$ matrix. The formula (\ref{entropy-on-shell}), together with the bound
(\ref{bound*}) that it suggests for the K{\"a}hler potential, is our main result.

\section{Summary}

We have obtained that the entropy of a general BH of type $(g,h)$ is given by 
(\ref{entropy-on-shell}). In words, the entropy is given by the sum of half
of Bekenstein-Hawking entropies for every horizon, plus the logarithm of the partition function.
In the case of the BTZ BH the logarithm of the partition function equals to the half of
the Bekenstein-Hawking entropy, which gives the entropy equal to the Bekenstein-Hawking value.
For a general BH the entropy is not given by a simple sum of horizon circumferences.
Indeed, the function $\ln{Z}$, which has the meaning of the K{\"a}hler potential on the Schottky space
(moduli space of BH's) depends non-trivially on the geometry of the BH interior. At best, the
entropy can be expected to be bounded from above by the Bekenstein-Hawking value, which leads
us to conjecture a very interesting bound (\ref{bound*}) on the K{\"a}hler potential. It would be
of considerable interest to establish this bound by pure mathematical means, without any reference to BH
physics. It would also be of interest to find if (and when) the bound can be saturated. In other words,
are there any other BH's apart from the BTZ whose entropy is equal to the sum of horizon circumferences over
$4G$? We leave these interesting questions to future work.

We conclude with a brief discussion of the physical interpretation of the entropy (\ref{entropy-on-shell}) 
we obtained. Let us first note that the analytic continuation procedure that we used was non-standard.
So is the entropy result (\ref{entropy-on-shell}) that we obtained.
Indeed, instead of analytically continuing the discrete group of identifications, we could have 
continued the time coordinate. Even though there is no global KVF whose affine parameter can serve as
a time coordinate covering the whole
spacetime (such a KVF exists only in the BTZ BH case), there is such a coordinate in every asymptotic region.
Indeed, every such region is indistinguishable from that of the BTZ BH. Thus, one can 
analytically continue the metric in some asymptotic region. The resulting space would be 
a solid torus, whose modular parameter is determined by the horizon size of the continued 
asymptotic region. The entropy in {\it that} asymptotic region would be equal 
to the corresponding horizon circumference over $4G$. One can argue that this is the entropy that is
``observable'' by an observer that lives in that asymptotic region. In contrast, the entropy obtained using  our 
analytic continuation procedure is that associated to the whole spacetime, not just 
a single horizon or a single asymptotic region. As such it is not a quantity that can
be detected or measured by an observer in a single asymptotic region. Instead, this is what a {\it super-observer}
that knows about the existence of all asymptotic regions (and about the topology inside the BH) 
would call the BH entropy. The entropy (\ref{entropy-on-shell}) is that for a BH of a particular 
internal geometry, and it depends on this geometry non-trivially.

\noindent
{\large \bf Acknowledgments}

I would like to thank D.\ Brill, S.\ Gukov, T.\ Jacobson, L.\ Takhtajan and P.\ Zograf for 
interesting discussions about this work. I am grateful to the IAS, Princeton for hospitality
during the time that this paper was written.

\end{document}